\documentclass[aps,floatfix,showpacs,prl,twocolumn]{revtex4}
\usepackage{graphicx,bm}

\begin{document}
\title{Long-lived quantum memory with nuclear atomic spins}

\author{A. Dantan${}^1$}
\author{G. Reinaudi${}^2$}
\author{A. Sinatra${}^2$}
\email{alice.sinatra@lkb.ens.fr}
\author{F. Lalo\"e${}^2$}
\author{E. Giacobino${}^1$}
\author{M. Pinard${}^1$}

\affiliation{${}^1$LKB, UPMC,
case 74, place Jussieu, 75252 Paris Cedex 05, France \\
${}^2$ \small LKB, ENS,
 24 Rue Lhomond, 75231 Paris Cedex 05, France}

\begin{abstract}
We propose to store non-classical states of light into the
macroscopic collective nuclear spin ($10^{18}$ atoms) of a $^3$He
vapor, using metastability exchange collisions. These collisions,
commonly used to transfer orientation from the metastable state
$2^{3}S_1$ to the ground state of $^3$He, can also transfer quantum
correlations. This gives a possible experimental scheme to map a
squeezed vacuum field state onto a nuclear spin state with very long
storage times (hours).
\end{abstract}


\maketitle

If great progress has been made in the generation of non-classical
states of light, a major challenge of quantum information and
communication lies in the ability to manipulate and reversibly store
such quantum states \cite{zoller,lukin}. Several proposals have been
made to achieve storage of non-classical light states either in
trapped cold atoms or atomic vapors \cite{fleischhauer,dantan1}. The
first successful experiments of quantum memories for coherent states
and squeezed states were achieved using atoms as a storage medium
\cite{polzik,manips_memoires}. In all the proposed schemes, as well
as in the experiments realized so far, the information is encoded in
the ground state of alkali atoms; the obtained storage times are at
most several milliseconds, limited by collisions, magnetic field
inhomogeneities, transit time, etc. Nuclear spins have also been
proposed as quantum memories for mesoscopic systems, due to their
long relaxation time \cite{lukin2}. In this Letter we show how to
reversibly map a non classical state of light into a squeezed state,
encoded in the purely nuclear spin of the ground state of ${}^3$He,
which interacts very little with the environment. The quantum state
can then survive for times as long as {\it hours}. To access the
ground state of ${}^3$He, which is 20 eV apart from the nearest
excited state, we propose to use metastability exchange (ME)
collisions, during which an atom in the ground state and an atom in
the metastable triplet state 2${}^3S_{1}$ exchange their electronic
variables. ME collisions are used in optical pumping of ${}^3$He to
create nuclear polarization in gas samples for nuclear physics
experiments as well as in nuclear magnetic resonance imaging
applications \cite{LeducReview}. When the helium vapor is in a
sealed cell, a weak radio-frequency discharge excited by a pair of
external electrodes maintains a tiny fraction of the atoms in the
metastable state, which has a finite lifetime due to its
interactions with the cell walls. A transition is accessible from
the metastable state to couple the metastable atoms with light.
This, together with ME collisions, provides an effective coupling
between the ground state atoms and light. We show that, with such a
mechanism, quantum fluctuations can be reversibly transferred from
the field to the atoms. Interacting with squeezed light in
appropriate conditions, the macroscopic nuclear spin ($1.6 \times
10^{18}$ atoms of ${}^3$He at 1 torr in a 50 cm${}^3$ cell, at 300
K) of the polarized ground state gas becomes squeezed. The nuclear
coherence relaxation time in absence of discharge and in an
homogeneous field can be several hours. By switching on the
discharge and repopulating the metastable state, the squeezing can
be transferred back to the electromagnetic field and measured. In
addition to its interest for quantum information, the scheme offers
the exciting possibility to create a long-lived non classical state
for spins.

\begin{figure}[htb]
\centerline{\includegraphics[width=8.5cm,clip=]{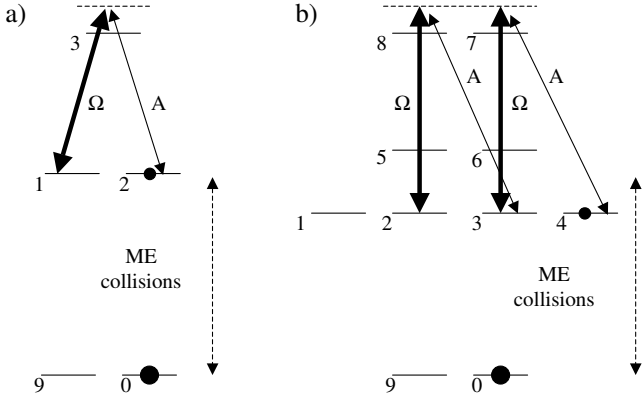}}
\caption{a) Sublevels $1$ and $2$ are metastable, level $3$ is the
excited state, $9$ and $0$ are the ground state sublevels. b)
Relevant energy levels in ${}^3$He.} \label{fig:schema}
\end{figure}

We consider a system composed by $N$ atoms in the ground state, and
$n$ atoms in the metastable state. These atoms interact with a
coherent driving field with Rabi frequency $\Omega$ and frequency
$\omega_1$ that we treat classically, and a cavity field described
by creation and annihilation operators $A$ and $A^\dagger$ (Fig.
\ref{fig:schema}-a). The field injected into the ring cavity,
$A_{in}$ with frequency $\omega_2$, is in an amplitude-squeezed
vacuum state: $\langle A_{in}\rangle=0$ and $\Delta
X_{in}^2=e^{-2r}$, $\Delta Y_{in}^2=e^{2r}$, where $X=A+A^{\dagger}$
and $Y=i(A^{\dagger}-A)$ are the standard field amplitude and phase
quadrature operators, satisfying $[X,Y]=2i$. The Hamiltonian of the
atom-field system is:
\begin{equation}
H=H_0 + \hbar \left\{ \Omega \, S_{31}e^{-i \omega_1 t} +
 g \, A  \, S_{32} + \rm{h.c.} \right\}
\end{equation}
where $H_0$ describes the free evolution of the atoms and the field,
$g=d (2\pi \omega_2 /\hbar V )^{1/2}$ is the coupling constant
between the atoms and the cavity field, $V$ being the volume of the
cavity mode, $d$ the atomic dipole. $S_{kl}= \sum_{i=1}^n
|k\rangle_i \langle l|_i$ for $k,l=1,2,3$ are collective atomic
operators in the metastable and excited state \cite{vecteurdonde}.
The coupling to the ground state collective spin
$I_{kl}=\sum_{i=1}^N |k\rangle_i \langle l|_i$ for $k,l=9,0$ is
provided by ME collisions.

We start with a simplified picture, in which both the metastable and
ground state atoms have a spin 1/2, which are simply exchanged
during each ME collision. The exchange collisions rate for a
metastable and a ground state atom are denoted by $\gamma_{m}$ and
$\gamma_{f}$, respectively. Their ratio $\gamma_{m}/\gamma_{f}$ is
equal to the ratio $N/n$. We assume that the system is initially
prepared using optical pumping in the fully polarized state $\langle
I_{00} \rangle=N$ and $\langle S_{22} \rangle=n$. Both the
metastable and ground state collective spins are polarized along the
$z$-axis of the Bloch sphere. The transverse spin components
$S_x=(S_{21}+S_{21}^\dagger)/2$, $S_y=i(S_{21}^\dagger-S_{21})/2$
then play a similar role to field quadratures and, for such a
coherent spin state, have equal variances: $\Delta S_x^2 = \Delta
S_y^2 = {n/4}$ and $\Delta I_x^2= \Delta I_y^2 = {N/4}$. By
definition \cite{ueda} the metastable (ground state) spin is
squeezed if one of the transverse spin variance $\Delta S_x^2$ or
$\Delta S_y^2$ ($\Delta I_x^2$ or $\Delta I_y^2$) is smaller than
its coherent spin state value. As usual in quantum optics, we study
the quantum fluctuations of operators around a ``classical steady
state" of the system (the fully polarized state). We then linearize
the equations, and obtain in the rotating frame the following closed
set of equations:
\begin{eqnarray}
\dot{{S}}_{21} &=& - (\gamma_{m}-i\delta) {S}_{21}+ \gamma_f
{I}_{09} - i {\Omega} {S}_{23} +
        f_{21} \label{eq:2NS} \\
\dot{{S}}_{23} &=& -(\gamma +i \Delta) {S}_{23} -
        i {\Omega} {S}_{21} - i gn  A +
        f_{23} \label{eq:2NPS}\\
\dot{{I}}_{09} &=& -(\gamma_{f}- i \delta_I) {I}_{09}+
\gamma_m {S}_{21} + f_{09} \label{eq:2NI} \\
\dot{A} &=& - (\kappa  + i \Delta_c) A -
        ig {S}_{23} + \sqrt{2\kappa } A_{in} \label{eq:2NA}
\end{eqnarray}
We have introduced the detunings $\delta=\omega_S-\delta_{las}$,
$\delta_I=\omega_I-\delta_{las}$, $\Delta=(E_3-E_2)/\hbar-\omega_2$
with $E_i$ the energy of level $i$, $\omega_I=(E_0-E_9)/\hbar$,
$\omega_S=(E_2-E_1)/\hbar$, $\delta_{las}=\omega_1-\omega_2$, and
the cavity detuning $\Delta_c=\omega_c-\omega_2$. $\Omega$ is
assumed to be real. The stochastic part of the evolution (quantum
noise) of each operator is described by a time-dependent Langevin
operator. If $\alpha$ and $\beta$ denote two system operators,
$\langle f_\alpha(t) f_\beta(t^\prime) \rangle = D_{\alpha \beta }
\delta(t-t^\prime)$ where $D_{\alpha,\beta}$ is the corresponding
coefficient of the diffusion matrix. Contributions to $D$ come from
polarization decay with a rate $\gamma$, ME collisions for
metastable and ground state atoms, and cavity losses with a rate
$\kappa$ for the cavity field. The non-zero coefficients of the
atomic part of the diffusion matrix are $D_{21,12}=D_{09,90}=2 n
\gamma_m$, $D_{21,90}=D_{09,12}=-2 n \gamma_m$, $D_{23,32}=2 n
\gamma$, calculated using the generalized Einstein relation
\cite{Cohen} for an ensemble of uncorrelated atoms. The Langevin
forces for ME collisions are necessary for the model to be
consistent. Otherwise the non Hamiltonian character of the exchange
terms leads to violation of the Heisenberg uncertainty relations.
Physically these forces originate from the fluctuating character of
the ME collisions. Their correlation time is the collision time,
much shorter than all the times scales we are interested in.

By adiabatic elimination of the polarization $S_{23}$ and the cavity
field assuming $\gamma, \kappa \gg \gamma_m, \gamma_f$, one obtains
\begin{eqnarray}
\dot{{S}}_{21} + (\gamma_m + \Gamma  -i \tilde{\delta} ) {S}_{21} &
= &
\gamma_f {I}_{09} + f_{21} \nonumber \\
-\frac{\Omega}{\Delta} f_{23}
    & + & i \frac{\Omega g n }{\Delta} \sqrt{\frac{2}{\kappa}} A_{in}
\label{eq:adiabS21}
\end{eqnarray}
where we introduced the optical pumping parameter
$\Gamma=\gamma{\Omega}^2(1+C)/\Delta^2$, and the cooperativity
$C=g^2 n/(\kappa \gamma)$, and we redefined the two-photon detuning
$\tilde{\delta}=\delta+\Omega^2/\Delta$ to account for the
light-shift of level 1. In deriving (\ref{eq:adiabS21}) we assumed a
Raman configuration $\Delta \gg \gamma$, $\frac{C\gamma}{\Delta}\ll
1$ and that the cavity detuning exactly compensates the cavity field
dephasing due to the atoms: $\Delta_C=C\kappa \gamma/\Delta$.
Optimal coupling between the squeezed field and the metastable
coherence is achieved under resonant conditions
$\tilde{\delta}=0$,~or
\begin{equation}
\omega_S(B)+\Omega^2/\Delta=\omega_1-\omega_2 \label{eq:matchmet}
\end{equation}
where the Larmor frequency $\omega_S$ can be adjusted using a
magnetic field. A second resonance condition is $\delta_I=0$,~or
\begin{equation}
\omega_I(B)=\omega_1-\omega_2 \, \label{eq:matchfond}
\end{equation}
meaning that the natural evolution frequency of the ground state
coherence $I_{09}$ should match that of the metastable state
coherence. The Larmor frequency in the metastable and ground states
are very different due to the difference between the nucleon and the
electron mass. In low field, $\hbar \omega_\alpha=\mu_\alpha B$
($\alpha$=I,S) with $\mu_I/h=3.24$kHz/G and $\mu_S/h=1.87$MHz/G.
However, the light shift in the metastable state allows to
simultaneously fulfill (\ref{eq:matchmet}) and (\ref{eq:matchfond})
for a non zero magnetic field. Physically, these conditions ensure
that both spin coherences are resonantly excited with the same
tunable frequency $\omega_I(B)$, thus ensuring the efficiency of the
squeezing transfer from the field to ground state. From Eq.
(\ref{eq:adiabS21}) and the corresponding equation for $I_{09}$ with
$\tilde{\delta}=\delta_I=0$, we can calculate the variances of the
metastable and ground state spins. In the limit $\gamma_f \ll
\Gamma, \gamma_m$ one obtains:
\begin{eqnarray} \Delta I_y^2 &=& \frac{N}{4} \left\{
1 - \frac{\gamma_m}{\Gamma+\gamma_m}
    \frac{C}{C+1} (1-e^{-2r}) \right\}
        \label{eq:varI}\\
\Delta S_y^2 &=& \frac{n}{4} \left\{ 1 -
\frac{\Gamma}{\Gamma+\gamma_m}
    \frac{C}{C+1} (1-e^{-2r}) \right\} \,.
        \label{eq:varS}
\end{eqnarray}
In the limit $C\gg1$, the squeezing can be completely transferred to
the atoms. If $\Gamma \gg \gamma_m$, correlations are established
among the metastable-state spins, the leakage of correlation towards
the ground state being negligible. The metastable collective spin is
squeezed while the ground state spin remains unsqueezed. In the
opposite limit $\Gamma \ll \gamma_m$, spin exchange is the dominant
process for metastable atoms; they transfer their correlations to
the ground state which then becomes squeezed. In all regimes the
metastable and the fundamental state share the amount of noise
reduction in the field.

In usual optical pumping experiments, the relevant atomic
observables are the level orientations, i.e. one-body observables.
ME collisions constantly tend to equal the degree of polarization of
the two levels. By contrast, squeezing a spin component amounts to
giving a negative value to the two-spin correlation function
$\langle s_y(1) s_y(2) \rangle$. ME collisions constantly tend to
equal the spin correlation functions in the two levels but not the
degree of squeezing. This is because, to create maximum squeezing, a
much smaller (absolute) value of the correlation function is needed
in the fundamental than in the metastable state, due to the large
population difference in the two levels \cite{noteFranck}. Somehow
paradoxically, the squeezed field then maintains a strong squeezing
in the ground state via a weakly squeezed metastable state.

If one switches on the discharge and the coherent field in the same
configuration as for the ``writing" phase \cite{dantan1}, the
nuclear spin memory can be ``read", the squeezing being transferred
back to the electromagnetic field where it can be measured. During
this process, the metastable level acquires only a weak degree of
squeezing under the effect of ME collisions. But, because of the
optical coupling, this squeezing progressively transits back to the
quantum field stored in the cavity, so that, in the end, a strong
squeezing is accumulated in the field without ever being large in
the metastable state.

One important issue is the ``writing" (or ``reading") time of the
quantum memory, which is the ground state effective response time.
The adiabatic elimination of the metastable state in Eq.
(\ref{eq:2NI}) shows that this time is the inverse of
$\Gamma_F=\frac{\gamma_f \Gamma}{\gamma_m + \Gamma}$, which is also
the width of the squeezing spectrum in the ground state.

\begin{figure}[htb]
\centerline{\includegraphics[width=7cm,clip=]{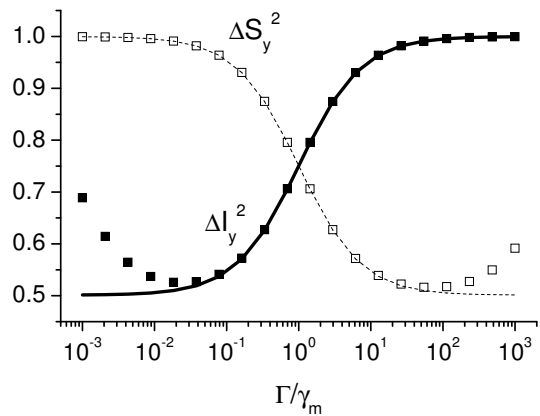}}
\caption{Analytical predictions (lines) and numerical calculations
for spin variances in ground state (full symbols) and metastable
state (open symbols), as a function of the ratio
$\Gamma/\gamma_{m}$. Numerical values of parameters are
$e^{-2r}=0.5$, $C=500$, $\kappa=100\gamma$, $\Delta=-2000 \gamma$,
$\gamma=2\times10^7$s$^{-1}$, $\gamma_m=5\times10^6$s$^{-1}$,
$\gamma_0=10^{3}$s$^{-1}$. } \label{fig:gamma0}
\end{figure}

We now apply our scheme to ${}^3$He atoms in realistic conditions
(Fig. \ref{fig:schema}-b). The coherent field ($\pi$-polarized) and
the squeezed vacuum ($\sigma^-$-polarized) are tuned to the blue
side of the so called $C_9$ transition ($\lambda=1.08$ $\mu$m) from
the $F=3/2$ level of the $2^3S_1$ metastable state to the ${}^3P_0$
state, the highest in energy of the $2^3P$ multiplicity
\cite{footnote}. The system is initially prepared in the fully
polarized state, $\langle I_{00} \rangle=N$ and $\langle S_{44}
\rangle=n$, by preliminary optical pumping. The metastable state now
has two sublevels $F=3/2$ and the $F=1/2$. The effect of ME
collisions on the metastable and ground state density matrices
$\rho_m$ and $\rho_f$ can be written as \cite{JDR}:
\begin{eqnarray}
\dot{\rho_f} &=& \gamma_f (-\rho_f + {\mbox{Tr}}_e \rho_m) \nonumber\\
\dot{\rho_m} &=& \gamma_m
        (-\rho_m + \rho_f \otimes {\mbox{Tr}}_n \rho_m)
\nonumber
\end{eqnarray}
where $\mbox{Tr}_e$ and $\mbox{Tr}_n$ represent trace operations
over the electronic and nuclear variables. After elimination of
hyperfine coherences and linearization around the initially prepared
state, we obtain a set of 11 closed equations involving the ground
state coherence, the cavity field, 4 optical coherences, the excited
state coherence, and 4 $\Delta m_F=1$ coherences in the metastable
state. To account for the fact that metastable atoms are destroyed
as they reach the cell walls, we introduce a damping rate $\gamma_0$
of the metastable state coherences. Despite the more complicated
level structure, in the fully polarized limit considered here, the
squeezing transfer to the ground state comes exclusively from the
coherence $S_{34}$ which should be excited resonantly. By adiabatic
elimination of the field and optical coherences, for optimal
squeezing transfer conditions and in the limit $\gamma_f \ll \Gamma,
\gamma_m$, we worked out the same analytical expressions
(\ref{eq:varI}-\ref{eq:varS}) for the ground state and metastable
spin variances, within a scaling factor in the optical pumping
parameter
\begin{equation}
\Gamma=\gamma{3 \Omega}^2(1+C)/\Delta^2 \,, \label{eq:Gamma}
\end{equation}
with now $\Delta=(E_7-E_4)/\hbar-\omega_2$. In Fig. \ref{fig:gamma0}
we show the analytical predictions (\ref{eq:varI}-\ref{eq:varS}) and
a full numerical calculation for realistic experimental parameters:
a 1 torr vapor at 300 K, with $\gamma_m=5\times10^{6}s^{-1}$, and
$\gamma=2\times10^7s^{-1}$, and a metastable atom density of $3.2
\times 10^{10}$ atoms/cm$^3$ corresponding to a ratio $n/N=10^{-6}$.
The relaxation rate $\gamma_0$ is inversely proportional to the gas
pressure (at 1 torr $\gamma_0=10^3s^{-1}$).
Deviations from the analytical formulas are due to non adiabaticity
of the optical coherence with respect to metastable variables, which
affects the squeezing of metastable spin, and to a finite relaxation
rate in the metastable state $\gamma_0$, which affects the ground
state spin squeezing in the region $\Gamma \ll \gamma_m$. In this
figure the one-photon detuning $\Delta$ is kept fixed while the
magnetic field and $\delta_{las}$ are chosen to satisfy
simultaneously (\ref{eq:matchmet}) and (\ref{eq:matchfond}) with now
$\omega_S=(E_4-E_3)/\hbar$. The energy positions of atomic levels in
the metastable and excited state depending on the field were
computed including the effect of hyperfine interactions
\cite{Zeeman}.

We calculated by simulation the effect of a frequency mismatch in
(\ref{eq:matchmet}) or (\ref{eq:matchfond}), on spin squeezing in
the ground state. A frequency mismatch of the order of $\Gamma/3$ in
the metastable state or of the order of $\Gamma_F$ in the ground
state affects the efficiency of the squeezing transfer. The
condition for the ground state frequency matching
(\ref{eq:matchfond}) imposes stringent requirements on the
homogeneity of the magnetic field \cite{champ}. Physically, if a
significant dephasing between the squeezed field and the ground
state coherence builds up during the squeezing transfer time, the
atoms will see an average between the squeezed and the antisqueezed
quadrature of the field, always above the standard quantum noise
limit. Let $\Delta B$ be the maximum field difference with respect
to the optimal value in the cell volume. For low field, the
condition on $\Delta B$ to preserve the transfer efficiency reads
$\mu_I \Delta B < h \Gamma_F$. Since $\Omega^2/\Delta \simeq \Gamma
\frac{\Delta}{3 \gamma C} \simeq \frac{\mu_S}{h} B$ we get
$\frac{\Gamma}{\Gamma_F}\frac{\mu_I}{\mu_S}\frac{\Delta}{3 \gamma C}
\frac{\Delta B}{B}<1$ or, in the regime $\Gamma \ll \gamma_m$, $600
\frac{\Delta}{\gamma C} \frac{\Delta B}{B}<1$.
\begin{figure}[htb]
\centerline{\includegraphics[width=8cm,clip=]{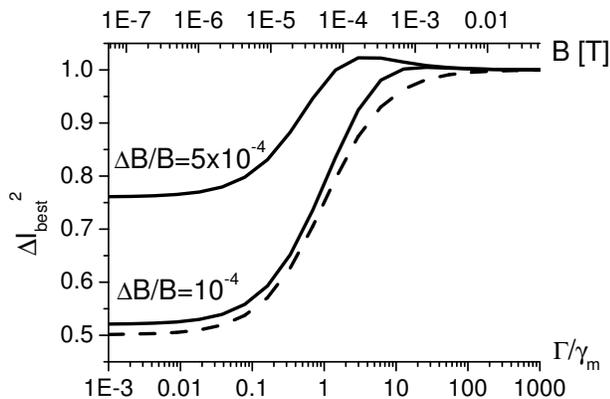}}
\caption{Ground state spin quadrature optimized (best) variance as a
function of the ratio $\Gamma/\gamma_{m}$ (lower $x$-axis), for two
relative changes of the magnetic field with respect to its optimal
value (upper $x$-axis). $\Delta B/B=0$ for the dashed curve.
$\gamma_0=0$ and other parameters are as in
figure~\ref{fig:gamma0}.} \label{fig:inrel}
\end{figure}
In Fig. \ref{fig:inrel} we show the effect of a relative change of
the magnetic field with respect to the optimal calculated value. An
homogeneity of 100 ppm is sufficient for the chosen parameters to
guarantee that all atoms are squeezed. The optimal calculated value
for the field is shown as a second $x$-axis in the figure. In
realistic conditions, choosing $\Gamma=0.1\gamma_m$, the required
field is about $B=57$mG, corresponding to $\omega_I=184$Hz. Squeezed
vacuum states that can be generated for analysis frequencies as low
as 200 Hz \cite{squeezing}, could thus be efficiently transferred to
the nuclear spins. The readout time is as long as the writing time:
$\Gamma_F^{-1}=2$s for $\Gamma=0.1\gamma_m$, limited by the
metastable atoms density in the sample.

The possibility to manipulate the spins using nuclear magnetic
resonance techniques, and to optically readout the spin state after
a long storage time makes this system particularly promising for
quantum information \cite{teleport}.

We thank Jacques Dupont-Roc, Nicolas Treps, Yvan Castin, Pierre-Jean
Nacher and Xavier Ma\^itre for useful discussions. Laboratoire
Kastler Brossel is UMR 8552 du CNRS de l'ENS et de l'UPMC. This work
was supported by the COVAQIAL European Project No. FP6-511004.


\begin{thebibliography}{99}
\bibitem{vecteurdonde}
Position dependent phase factors due to field propagation are
ignored since we assume that the fields are copropagating, far from
resonance, and with very small frequency difference in the working
conditions of the memory.
\bibitem{noteFranck}
The two spin correlation function is subject to an absolute lower
limit that is proportional to the inverse total number of spins in
the level, due to a simple positivity condition. When this limit is
reached, the corresponding component of the total spin is completely
squeezed. In our calculation, the spin correlation functions in the
two levels become both equal to $(\exp(-2r)-1)/4(n+N)$ in the
extreme limit $\Gamma \le \gamma_f$.
\bibitem{footnote}
We have assumed above that the squeezed field and the coherent field
are copropagating. The squeezed field should thus be
$\sigma$-polarized, with a $\sigma^+$ component; we checked that
this component plays no role here.
\bibitem{champ} Although the Larmor precession in the metastable
state is $\sim 10^3$ times faster than in the ground state, the
effective transfer time (interaction with the field and
metastability exchange) is $\sim 10^{6}$ times shorter for the
metastable state, so that the most stringent condition on $B$ comes
from the ground state.
\end{thebibliography}
\end{document}